\documentclass[reprint,amsmath,amssymb,aps,prb]{revtex4-2}

\usepackage{graphicx}
\usepackage{dcolumn}
\usepackage{bm}
\usepackage{ amssymb }
\usepackage{graphicx}
\usepackage{dcolumn}
\usepackage{color}
\usepackage{bm}
\usepackage{url} 
\usepackage{hyperref}
\hypersetup{
 colorlinks   = true,
 linkcolor=blue,
  citecolor    = blue,
  urlcolor=blue,
  } 

\usepackage{physics}
\usepackage[normalem]{ulem}
\usepackage[dvipsnames]{xcolor}
\newcommand{\outstrike}[1]{}

\usepackage{enumitem}   
\begin{document}
\title{Accurate measurement of energy relaxation via flux-flow instability}

\author{E.M. Baeva$^{1,2}$, N.A. Titova$^{1}$, M.A. Kirsanova$^{3}$, S.A. Evlashin$^{3}$,\\ A.V. Semenov$^{1,4}$, D. Yu. Vodolazov$^{5}$, A.I. Kolbatova$^{1}$, and  G.N. Goltsman$^{2,6}$}

\affiliation{$^1$Moscow Pedagogical State University, Moscow, Russia\\$^2$HSE University, Moscow, Russia\\
$^3$Skolkovo Institute of Science and Technology, Moscow, Russia\\
$^4$Moscow Institute of Physics and Technology, Moscow, Russia\\
$^5$Institute for Physics of Microstructures, RAS, Nizhny Novgorod, Russia\\
$^6$Russian Quantum Center, Moscow, Russia
}

\begin{abstract}

The flux-flow instability (FFI) technique is a widely used method for measuring the quasiparticle relaxation time ($\tau_E$) in superconductors. In this study, we investigate the established FFI models by using a single-crystal superconducting titanium nitride (TiN) film with negligible pinning and slow electron-phonon relaxation. We study the critical current density of 12-nm-thick TiN samples with varying widths and find that the vortex velocity and $\tau_E$ can be precisely determined when the strip width exceeds the electron-phonon relaxation length. Comparative analysis of the energy relaxation times obtained through FFI and other methods reveals that the relaxation of quasiparticles within the vortex core is predominantly driven by an increase in quasiparticle temperature relative to phonons, contrary to the predictions of the Larkin-Ovchinnikov and Bezuglyi-Shklovskii models. In contrast, the Kunchur model accurately describes the experimental data for TiN samples. The analysis suggests that $\tau_E$ can be reliably determined using the FFI method, with careful consideration for experimental conditions and material parameters.

\end{abstract}

\maketitle
\section{INTRODUCTION}

The study of vortex dynamics in superconducting devices is an essential topic in both fundamental research and applied superconductivity~\cite{Dobrovolskiy2024}. In type-II superconductors in a magnetic field, vortices initially move by creep as the current increases. At higher currents, the system enters the flux-flow regime, where the vortices move with a macroscopic velocity exiting quasiparticles. When the velocity reaches its maximum $v=v^*$, the quasiparticles cannot dissipate, and flux-flow instability occurs. Maximizing $v^*$ is crucial for applications such as ultrasound generation \cite{Bulaevskii2005,Ivlev1999}, magnon excitation \cite{Bespalov2014,DobrovolskiyCherenkov}, and microwave radiation \cite{Dobrovolskiy2018Microwave,Lsch2019}.  Moreover, studying $v^*$, provides valuable insights into the energy relaxation time, $\tau_E$, which governs the performance of superconducting devices. When the vortex passing time becomes less than $\tau_E$, quasiparticles within the vortex core are unable to relax, leading to flux-flow instability (FFI). There are two different models that describe this phenomenon in weakly pinned superconductors. The Larkin-Ovchinnikov (LO) model assumes a non-thermal quasiparticle distribution \cite{LO_nonlinear}, where energy dissipation occurs through quasiparticles escaping the vortex core. In contrast, the Kunchur model assumes that the quasiparticle distribution has reached thermal equilibrium, and energy relaxation arises from the heating of quasiparticles \cite{Kunchur2002}. These two approaches provide different descriptions of the FFI behavior depending on the experimental conditions, such as temperature, magnetic field, and the quasiparticle distribution function.

The LO approach has recently been shown to be an effective method for characterizing materials used in superconducting single-photon detectors \cite{Lin2013,Hofer2021,Liu2021,Caputo2017,Cirillo2021,Dobrovolskiy2017,Bevz_2023,Yadav2024}. By measuring the critical vortex velocity, $v^*$, one can determine the energy relaxation time $\tau_E$ using the LO model. This information allows to describe the hot spot dynamics following photon absorption in these detectors. However, there are significant discrepancies between $\tau_E$ values obtained using the vortex velocimetry method and those obtained from other experimental techniques, such as photoresponse and magnetoconductance measurements \cite{Gershenzon1990,Lin2002}. These discrepancies can be observed in various materials, including NbN, NbGe, TiN, and NbC. In some cases, $\tau_E$ value obtained with the LO method is overestimated (e.g., NbN and NbGe) \cite{Yadav2024,Lin2013}, while in others it is underestimated (e.g., TiN, NbC, NbN) \cite{Lefloch1999,Dobrovolskiy2020_NBC,Haberkorn2024}. Furthermore, the experimentally observed exponential temperature dependence of $\tau_E$ is inconsistent with the expected $T^{-3}$ scaling for electron-phonon (e-ph) relaxation \cite{Haberkorn2024,Zhang2020,Leo2011,Ulacco2025}. Additionally, superconducting single-photon detectors typically employ low-$T_c$ \cite{Tripathy2024} and strongly disordered \cite{Zolotov2021} superconducting films, where strong electron-electron (e-e) scattering can result in a thermal-like distribution of quasiparticles \cite{Giazotto2006,Kozub1995, Nagaev1995}. This scenario is better explained by the Kunchur model rather than the LO approach. Therefore, it is essential to carefully consider both experimental conditions and material properties when interpreting the results of vortex velocity measurements.

To emphasize the critical role of quasiparticle heating effects, we investigate the flux-flow instability (FFI) in a 12 nm thick TiN single-crystalline film with negligible volume pinning. This material is particularly valuable, as its e-ph relaxation time, $\tau_{eph}$, has been measured directly \cite{Baeva2024_noise}. In TiN the e-ph and e-e times are similar ($\tau_{eph}\sim\tau_{ee}$) near $T_c$, making it difficult to select an appropriate theoretical model for interpreting FFI results a priori. We measured the critical current density in TiN strips with varying widths to identify the device parameters suitable for FFI measurements. Depending on the theoretical approach employed, we obtained different estimates for $\tau_E$, which do not directly correspond to $\tau_{eph}$ measurements. To better fit the data, we introduce a coefficient into the Kunchur model to account  the transition from one-dimensional to two-dimensional diffusion of quasiparticles. This modification leads to excellent agreement with the experimental data, as the extracted $\tau_E$ closely corresponds to $\tau_{eph}$. This consistency strongly supports the Kunchur model for vortex core energy dissipation. Thus, our findings demonstrate that $\tau_E$ below $T_c$ can be accurately determined from FFI measurements.

\section{Probing the energy relaxation time via vortex velocimetry}

\begin{figure}[h!]
    \centering
 \includegraphics[width=0.48\textwidth]{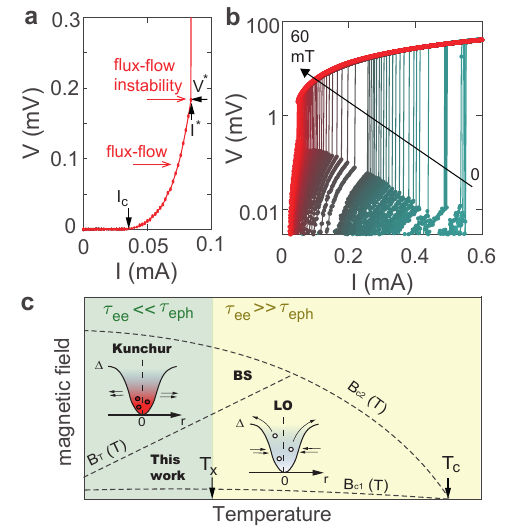}
\caption{a) The typical IV curve for a representative TiN sample. The data are presented for sample A5 (\autoref{Table1}) at $T$ = 4.2 K and $B$ = 25 mT on a linear scale with black arrows indicating $I_c$, $I^*$, and $V^*$. b) A set of $IV$-curves for sample A5 at different values of the perpendicular magnetic field $B$, at $T$ =4.2 K, on a log-log scale. c) An illustration of different conditions for the FFI in the $B(T)$ phase diagram. The mixed state for a type-II superconductor, where vortices exist, is located between the first and second critical magnetic fields ($B_{c1}(T)$ and $B_{c2}(T)$). The FFI regimes are separated by the crossover temperature $T_x$, at which $\tau_{ee}=\tau_{eph}$, and the overheating field $B_T(T)$, as described by the BS model.}
 \label{figure_num_0}
\end{figure}

In type-II superconductors, when an external magnetic field and a bias current are applied, the vortex lattice begins to move with a velocity $v$ due to the action of the Lorentz force, $\Phi_0 v$. This flux-flow is opposed by viscous forces $\eta v$, where $\eta$ is the viscosity coefficient. As a result, a voltage $V=-\left[v \times B\right] L $ appears along the superconductor. As $v$ increases, the viscous force also increases until it reaches its maximum value at the critical vortex velocity $v^*$. At this point, the FFI occurs, disrupting the superconducting state. \autoref{figure_num_0}(a) shows the current-voltage (IV) curve of a typical TiN sample under a constant magnetic field. Above the critical current $I_c$, the voltage increases monotonically with the applied current until a sharp increase occurs at an instability current $I^*$, corresponding to the onset of the FFI. The threshold voltage $V^*$ at the instability current $I^*$ corresponds to the state where the vortex velocity reaches its maximum: $v^*=V^*/(L B)$. 

\autoref{figure_num_0}(b) illustrates how the IV curves evolve with the magnetic field. Analyzing the dependence of $V^*$ on $B$ allows us to determine $v^*(B)$ dependence. In the following sections, we will examine models that describe $v^*(B)$ and extract the energy relaxation time in the vortex core. \autoref{figure_num_0}(c) illustrates the conditions of applicability of different FFI models. When $T\lesssim T_c$ and the quasipartice energy distribution is considered to be non-thermal, the vortex velocity $v^*$ is described by the LO model \cite{LO_nonlinear}. When a voltage $V$ is applied to a superconductor, quasiparticles in the core of the vortex acquire an energy that is different from the energy in the equilibrium state and begin to diffuse away from the core. The diffusion of quasiparticles causes the vortex to compress, and as a result, the viscosity of the vortex lattice, which is related to the quasiparticle distribution function and the superconducting gap, decreases as $v$ increases. At the critical velocity $v^*$, the IV curve exhibits instability, i.e. a straight or N-shaped jump. The authors in ref.~\cite{Doettinger1995} showed that $v^*$ does not linearly depend on $B$, i.e. $v^*$ is not a constant value, and they introduced the condition $L_E\gg a$, where $\alpha_{LO}=\left(2 \Phi_0 /\sqrt{3}B\right)^{1/2}$ is the period of the triangular vortex lattice and $L_E=\sqrt{\tau_E D}$ is the energy relaxation length. Thus, the equation described $v^*(B)$ can be written as~\cite{Doettinger1995}:

\begin{equation}\label{eq:1}
  v^*=  \left[14\zeta(3) \left(1-T/T_c \right)\right]^{1/4}  \left[\frac{D}{\pi \tau_E }\right]^{1/2}\left[1+\frac{\alpha_{LO}}{\sqrt{D \tau_E}}\right]
\end{equation}

As was shown later by Bezuglyi and Shklovskii (BS)~\cite{Bezuglyj1992}, if the magnetic field exceeds a certain characteristic field, $B_T$, the FFI can no longer be treated independently of the heating effects caused by the viscous flux flow. Above $B_T$, the quasiparticle distribution function changes in a thermal manner, even when $\tau_{ee}\gg\tau_{eph}$. In the case of a high substrate thermal conductivity, low Kapitza resistance, and a strip length that is greater than the e-ph relaxation length ($L\gg L_{eph}$), the value of $B_T$ can be determened as $B_T=0.347 k_B^{-1}e R_s d G_{eph}(T)\tau_{eph}(T)\propto T$ ~\cite{Bezuglyj1992,Bezuglyj2019}, where $G_{eph}$ is the e-ph thermal conductance. At $B>B_T$, the vortex velocity is determined by ~\cite{Bezuglyj1992}:
\begin{equation}\label{eq:2}
  v^*=  \left[\frac{G_{E} T_c \rho_n}{2.02 B B_{c2}(0)}  \right]^{1/2} \left[\frac{T_c-T}{3 T_c} \right]^{1/4}.
\end{equation}
Here, $G_{eph}$ can be expressed as $G_{eph}=5(\Sigma_{eph}/\nu)T^4$ with the e-ph cooling rate $\Sigma_{eph}$ and the sample volume $\nu=dwL$ \cite{Huard2007}.

If $T\ll T_c$, and the quasipartice energy distribution is considered to be thermal-like, the vortex core will be broadened instead of compressed, and the vortex viscosity decreases due to a change in the vortex profile. Combining the Bardeen-Stephen assumption for flux-flow resistance and the heat balance equation, Kunchur model~\cite{Kunchur2002} predicts $v^*$ in the limit where $\Delta(T)\ll k_B T$ and $B>B_T$: 
\begin{equation}\label{eq:3}
  v^*= \left[\frac{ \rho_n}{ B B_{c2}(T_e)\tau_E} \int^{T_e}_{T_b} C_e dT \right]^{1/2}.
\end{equation}
Here, $T_e$ is the electron temperature calculated as $T_e=(T_b+I^*V^*/\nu\Sigma_{eph})^{1/5}$, $T_b$ is the bath temperature, $C_e=\frac{\pi^2}{3}N_0 k_B^2 T$ is the electron specific heat capacity, $N_0=(e^2D R_s d)^{-1}$ is the density of states at the Fermi level, and $B_{c2}(T_e)=4k_B(T_c-T_e)/\pi D e$ is the second critical field at $T=T_e$.

The crossover temperature $T_x$, shown in \autoref{figure_num_0}(c), marks the transition between a non-thermal and a thermal quasiparticle energy distribution functions. A thermal distribution occurs when e-e collisions are much faster than other relaxation processes, such as e-ph scattering, i.e. $\tau_{ee}\ll\tau_{eph}$. In this case, the quasiparticle system reaches equilibrium and the electron distribution function $f$ follows a Fermi-Dirac distribution with an effective electron temperature, $T_e$. Considering the known $T$-dependencies for $\tau_{ee}\propto T^{-3/2}$  \cite{Aleiner2002} and $\tau_{eph}\propto T^{-3}$ \cite{Huard2007}, the condition $\tau_{ee}\ll \tau_{eph}$ can be used to determine the crossover temperature $T_x$ in a specific material system.

To determine the temperature range of our experiment in relation to the crossover temperature, $T_x$, we compare the experimentally obtained e-ph relaxation time $\tau_{eph}$ with the calculated e-e energy relaxation time $\tau_{ee}$ for epitaxial TiN, and also compare the applied magnetic field $B$ with $B_T$. The noise thermometry and noise spectroscopy \cite{Baeva2024_noise} provide information on the e-ph cooling rate $\Sigma_{eph}=1.35\times 10^8$ WK$^{-5}$m$^{-3}$ and the e-ph time $\tau_{eph}(T)=340 T^{-3}$ ns for identical TiN films. The latter corresponds to $\tau_{eph}(T_c)=2.6$ ns. Next, we estimate the e-e energy relaxation time $\tau_{ee}$ according to Aleiner et al. \cite{Aleiner2002} $\tau_{ee}=\hbar N_0 (\hbar D/k_B T)^{3/2}$, where $N_0=(eR_sdD)^{-1}=65$ eV$^{-1}$nm$^{-3}$ is the density of states. From $\tau_{ee}(T)$ and $\tau_{eph}(T)$, we estimate the crossover temperature $T_x\simeq 6.6$ K. Taking into account $\Sigma_{eph}$, we also obtain $B_T\simeq 0.5$ T. Thus, in our measurements, $B\ll B_T$ and the overheating of quasiparticles does not exceed 0.1 K at $T=4.2$ K (see Appendix A for the details). The estimated parameters suggest that the samples may be in either in the LO regime ($B<B_T$, $\tau_{ee}\gg \tau_{eph}$) or an as-yet unexplored region ($B<B_T$, $\tau_{ee}\ll \tau_{eph}$). In the latter case, the Kunchur model may offer reasonable approximations. In the next section, we will systematically investigate all available models in order to determine which one predicts results consistent with the experimentally observed energy relaxation times in TiN.

\section{SAMPLES AND METHODS}
\begin{figure}[h!]
    \centering
    \includegraphics[width=0.48\textwidth]{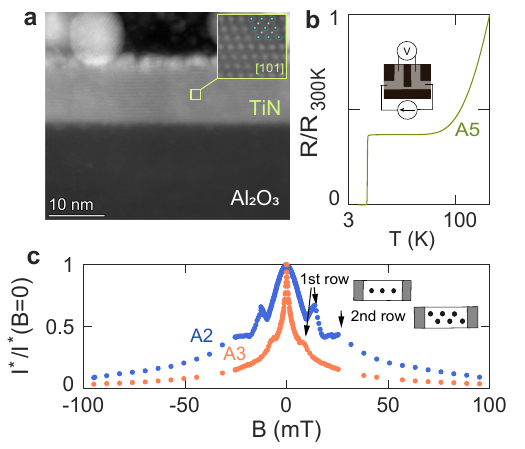}
    \caption{a) TEM cross-sectional images of TiN thin film deposited on Al$_2$O$_3$ substrate. Inset shows atomic packing in the [101]-oriented cubic structure of TiN. Blue and gray spheres of the superimposed projection of the structure correspond to Ti and N atomic columns, respectively. b) Main: Temperature dependence of normalized resistance of sample A5 plotted on a semi-log scale. Inset: An optical microscopy image of sample A5, and a schematic diagram of the experimental setup for current-voltage (I-V) curve measurements. c) Normalized critical current vs magnetic field for samples A2 and A3 at 4.2 K. The black arrows indicate the peaks, which correspond to vortex chains entering the strip, as illustrated in the sketches.}
    \label{figure_num_2}
\end{figure}

The samples studied here are patterned from a 12-nm-thick TiN film grown by DC reactive magnetron on a c-cut sapphire substrate. High-resolution TEM imaging reveals the single-crystalline structure of this TiN film (see \autoref{figure_num_2}(a)). We patterned the samples into strips using a combination of optical lithography, scanning electron-beam lithography, and plasma-chemical etching. \autoref{Table1} lists all the samples investigated in this study. Sample A1 is a two-contact strip measured in a quasi-four-probe configuration. Samples A2-A7 are Hall bars measured in a four-probe configuration. 

\begin{table}[h!] 
\begin{tabular}{ccccccccc}
\hline
\textnumero & $w$      & $L$      & $T_c$  & $j_c$& $w/\xi_{GL}$& $w/\Lambda_P$& $w/L_{eph}$\\
        & $\mu$m & $\mu$m & K & MA/cm$^2$ &  &   & \\ \hline \hline
A1       & 0.064    & 8        & 5.07   &1.74&1.2 &0.008 &0.03 \\ 
A2       & 0.5    & 22      & 5.1      &1.67& 9.5& 0.06&0.25\\ 
A3       & 1     & 23        & 5.11    &2.27& 19& 0.12&0.5\\ 
A4        & 3    & 26       & 5.11      &1.75 & 57 & 0.36&1.5\\ 
A5       & 5    & 36       & 5.1     &0.93&95 & 0.6&2.5\\ 
A6       & 10     & 44       & 5.11     &0.46 & 190& 1.2&5\\ 
A7        & 500    & 1000    & 5.12       &0.39& 9488& 60.3 &250\\ 
\hline
\end{tabular}
\caption{\label{Table1} Parameters of TiN samples.}
\end{table}

Resistance measurements presented in \autoref{figure_num_2}(b) are performed using AC Lake Shore 370 resistance bridge in a four-probe configuration and the bath temperature $T$ is monitored using a calibrated Lake Shore thermometer positioned near the sample. \autoref{figure_num_2}(b) shows $R(T)$ curve for a representative sample (A5). The critical temperature $T_\mathrm{c}$ is determined as temperature at which the sample lost half of its resistance, $R = R_\mathrm{n}/2$. The sheet resistance is estimated as $R_\mathrm{s} = R_n/N_\mathrm{s}=9$ $\Omega$/sq, where $R_n$ is normal-state resistance at 6\,K, $N_\mathrm{s} = L/w$ is the number of squares, $w$ and $L$ are the width and the length of samples. A magnetic field is applied perpendicular to the sample plane and varied up to 100 mT. The $IV$-curves are measured using a Keithley instrument set (Nanovoltmeter 2128A and 6221 AC Current Source). The parameters of all samples, including $w$, $L$, $T_c$, and $j_c= I_c$(4.2 $K$,$B=0$ T)$/(wd)$ are listed in \autoref{Table1}. Using the diffusion coefficient $D=8.5$ cm$^2$/s, extracted from the slope of the upper critical field $B_{c2}(T)$ \cite{Baeva2024_rt}, we estimate the depairing current density ($j_{dep}$) using the formula: $j_{dep}(T)= 0.74 \Delta_0^{3/2}(\hbar D)^{-1/2}(R_s e)^{-1}(1-(T/T_c)^2)^{3/2}$ \cite{KL1980}, where $\Delta_0=1.764 k T_c$ is the superconducting energy gap. The maximum experimental value of $j_c(4.2K)$ reaches 0.72$j_{dep}(4.2K)$ (sample A3), where $j_{dep}(4.2K)=3.154$ MA$^2$/cm$^2$. We also add in \autoref{Table1} the ratio of $w$ to the coherence length, $w/\xi_{GL}$, the ratio of $w$ to the Pearl length, $w/\Lambda_{P}$, at 4.2 K
\footnote{Note: The Ginzburg-Landau coherence length is calculated as $\xi_{GL}=\xi_{GL}(0)/\sqrt{(T_c-T)/T_c}$, where $\xi_{GL}(0)^2=\pi \hbar D/8k_B T_c$. The Pearl length is calculated as $\Lambda_{P}=2\lambda(T)^2/d$, where $\lambda(T)=\lambda_0 \sqrt{1-(T/T_c)^4}$\cite{Tinkham1995} is the penetration depth and 
$\lambda_0=\sqrt{\hbar R_s d/(\pi\Delta(0)\mu_0)}$=130 nm\cite{Kamlapure2010} is the zero-temperature penetration depth.}, and the ratio of $w$ to the e-ph length, $L_{eph}=\sqrt{\tau_{eph}D}$, at 4.2 K.

Analysis of the $IV$-curves can also provides information about quality of superconducting properties of the single-crystalline TiN film. \autoref{figure_num_2}(c) demonstrate a peak-shaped dependence of $I_c$ on the magnetic field observed for samples A2 and A3. The samples have widths of 10 and 20 $\xi_{GL}$. According to the Ginzburg-Landau model, in the case of weak volume pinning, vortex chains can alternate entering a strip at certain sample widths. This effect is called the peak effect \cite{Vodolazov2013,ichkitidze1981peak, Carapella2016}, since each entering chain manifests itself in the appearance of a minimum and maximum of the critical current. Thus, this peak effect characterizes the 12-nm TiN film as a film with extremely weak volume pinning. 

\section{Vortex velocimetry}

The data for the vortex velocity $v^*$ as a function of $B$ for all TiN devices studied in this work are shown in \autoref{figure_num_3}(a). As shown in \autoref{figure_num_3}(a), $v^*$ depends on both the magnetic field and the strip width. The non-monotonic $v^*(B)$ behavior at low fields are consistent with previous studies \cite{Shklovskij2017,Grimaldi2010}, but the increase in $v^*$ with $w$ in a finite magnetic field (\autoref{figure_num_3}(a)) is a novel finding. To better understand this phenomenon, we analyze the region where the flux-flow occurs (from $I_c$ to $I^*$, \autoref{figure_num_0}(a)). We normalize this region to $I^*$ and plot it as a function of $B$ in \autoref{figure_num_3}(b). For the widest samples (A6 and A7), the flux-flow region appears almost immediately and remains constant, while in intermediate-width samples (A4 and A5), the flux-flow region emerges at 3-4 mT and grows monotonically with $B$. Notably, the flux-flow region is entirely absent in the narrow samples A1-A3 (not shown in the figure). The simultaneous reduction in $v^*$ and the flux-flow region resembles the result of transition from uniform vortex motion to a region where vortices move at different speeds, similar to the effect of strong pinning \cite{Dobrovolskiy2024}. 

To better understand the experimental conditions for vortex velocimetry in TiN, it is essential to comprehend the factors that influence the critical current density ($j_c$). In order to study the FFI, it is necessary to have a sample width that meets specific criteria: $4.4\xi_{GL}<w<\Lambda_P$ (marked by the arrows in \autoref{figure_num_3}(b)). The lower boundary of the sample width at $B=0$ should be at least $4.4\xi_{GL}$, known as Likharev's limit \cite{Likharev1979}, where vortices start to emerge within the strip. The upper boundary must be less than the Pearl length $\Lambda_P$, ensuring a uniform supercurrent density at $B=0$. There are three distinct regions in $j_c(w)$ dependence in TiN strips:

\begin{figure}[h!]
    \centering
    \includegraphics[width=0.48\textwidth]{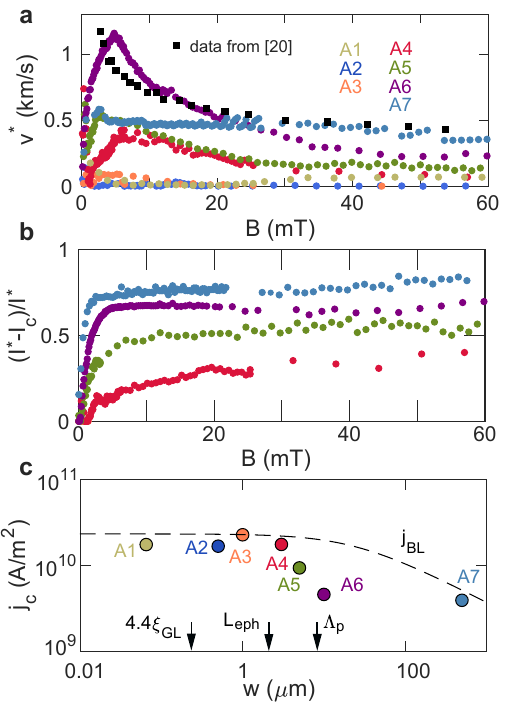}
    \caption{ a) The $B$-dependence of vortex velocities for samples A1-A7 measured at 4.2 K in comparison with previously reported data for TiN film \cite{Lefloch1999}. b) The $B$-dependence of current area for the flux-flow at 4.2 K. c) The critical current density $j_c$ is measured for samples A1-A7 measured at 4.2 K (symbols) in comparison with the following from edge barrier model critical current density $j_{BL}$ (dashed line). The arrows show the Likharev's limit $4.4\xi_{GL}(4.2$ K), the Pearl length $\Lambda_{P}(4.2$ K), and the electron-phonon length $L_{eph}(4.2$ K).}
    \label{figure_num_3}
\end{figure}

\begin{description}
\item[Region I] We observe higher values of $j_c$ for narrow strips (A1-A3). When the width of the samples approximately corresponds to $\sim 4.4\xi_{GL}$ = 230 nm in TiN (sample A1), as indicated by the arrow in \autoref{figure_num_3}(b), vortices are not expected to exist in this region. This allows $j_{dep}$ to be achieved \cite{Ilin2010}. 

\item[Region II] $j_c$ tends to decrease gradually as the strip width increases (A4-A6). Classically, this effect in thin films ($d\ll\lambda$) is due to nonuniform current distribution as width of the strip becomes comparable to or larger than the penetration depth $\Lambda$ \cite{Plourde2001}. This can be described by the formula \cite{Plourde2001}: $j_{BL}=j_s\sqrt{(\Lambda_p w/(1/\pi+\Lambda_p/w)}/w$, where $j_s$ approximately $0.75 j_{dep}$ for our case. However, our results show that $j_c(w)$ decreases more rapidly with increasing strip width than predicted. The origin of this effect remains unclear. While a similarly rapid decrease in $j_c$ has been attributed to the self-field effect reached the first critical field, $B_{c1}$, \cite{Ilin2010,Ilin2012,Engel2008}, we argue this cannot be the case. The standard calculation of $B_{c1}$ is invalid here, as it assumes a uniform perpendicular magnetic field. We instead speculate that the rapid decrease in $j_c$ could be attributed to the increasing area of the strip as its width $w$ grows. A larger area presents a higher probability of encountering a sizable defect somewhere within the strip (not necessarily near the edge), which would lead to a stronger suppression of the critical current $I_c$ and consequently $j_c$. An analogous effect, where $I_c$ decreases with the length of a superconducting meander/strip due to increased defect probability, has been discussed in \cite{Xu2023,Gaudio2014}.

\item[Region III]
For the widest samples (A6-A7), the critical current density $j_c$ reaches a minimum value, most probably related to the vortex pinning.
\end{description}

Based on our data, single-crystal TiN samples can be categorized by their vortex dynamics into narrow (A1-A3) and wide (A4-A7) strips. In narrow samples, the critical current density ($j_c$) is relatively large at small and intermediate magnetic fields. This leads to high vortex velocities even at the critical current ($I = I_c$), resulting in FFI and the absence of a resistive branch on the current-voltage IV-curve. In contrast, the $j_c$ in wide strips is strongly suppressed by the magnetic field, permitting a resistive state where we observe $I^*$ and $V^*$ (see \autoref{fig_sup1}(a) in the Appendix B).

A possible reason for the increase of the critical vortex velocity $v^*$ with strip width $w$ could be connected to the extremely large electron-phonon relaxation length ($L_{eph}$, see \autoref{Table1}), which originates from a large value of $D$ and $\tau_{eph}$. When $w < L_{eph}$, the cooling of nonequilibrium quasiparticles by diffusion is less effective because they can lose their energy only by moving along the strip (1D diffusion). In contrast, when $w > L_{eph}$, the diffusion becomes 2D, leading to a more effective cooling and a larger $v^*$. This geometric effect can make the measured flux-flow resistance ($R^*=V^*/I^*$) to deviate from the classic Bardeen-Stephen prediction ($R^*/R_n \simeq (1/\beta(T)) B/B_{c2}(T)$), where $\beta(T)=1.1 (1-T/T_c)^{-1/2}$ \cite{GOR_KOP1,Gorkov1975}. To account for the actual fraction of flux-flow resistance, we introduce a coefficient $\alpha$, which adjusts for the real fraction of flux-flow resistance $R^*$ (see the Appendix C for details). This leads to our modified Bardeen-Stephen relation: $R^*/R_n \simeq (\alpha/\beta(T)) B/B_{c2}(T)$, where $\alpha$ varies from 0 (for narrow samples $w<L_{eph}$) to 1 (for wide samples $w\gg L_{eph}$). Using this relation, we modify Eq. \eqref{eq:3} to the following form:
\begin{equation}\label{eq:4}
  v^*= \left[\frac{\alpha}{ \beta(T)}\frac{\rho_n}{ B B_{c2}(T_e) \tau_E} \int^{T_e}_{T_b} C_e dT \right]^{1/2}.
\end{equation}

\section{RESULTS AND DISCUSSION}

\begin{figure}[h]
    \centering
    \includegraphics[width=0.48\textwidth]{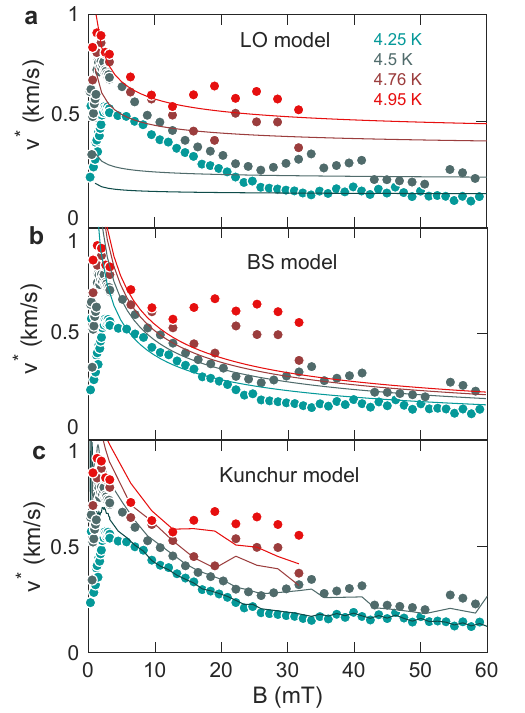}
    \caption{The vortex velocity $v^*$ as a function of the perpendicular magnetic field B at a set of different temperatures for sample A5 is compared to the $v^*$ values from the LO model(a), BS model(b), and Kunchur model(c).}
    \label{figure_num_4}
\end{figure}

We initially investigate which vortex motion models best describe our findings. \autoref{figure_num_4} compares the magnetic field ($B$) dependence of the critical vortex velocity ($v^*$) for sample A5 with predictions from three theoretical frameworks: the LO model, the BS model, and the Kunchur model.
Panel (a) shows the LO model fits (solid lines), derived from \autoref{eq:1}, using the experimentally obtained values of $D=8.5$ cm$^2$/s and $T_c=5.1$ K. Panel (b) presents the BS model fit, using \autoref{eq:2}, with the same values of $D$ and $T_c$, and $B_{c2}(0) = 0.66$ T. Here, we also used the electron-phonon cooling coefficient $\Sigma_{eph}=1.35\times 10^8$ WK$^{-5}$m$^{-3}$, derived from \cite{Baeva2024_noise}, to calculate the electron temperature $T_e=(T_b+I^*V^*/\nu\Sigma_{eph})^{1/5}$, and the value of $B_{c2}(T_e)$.
The Kunchur model, which is presented in panel (c), can be described using \autoref{eq:3} and \autoref{eq:4}. In these equations, $\tau_{E}$ and $\alpha$ are the only fitting parameters. Although the LO and BS model show some discrepancies compared to the experimental data, our analysis suggests that Kunchur model provides more consistent fits to the data.

\begin{figure}[h!]
    \centering
    \includegraphics[width=0.48\textwidth]{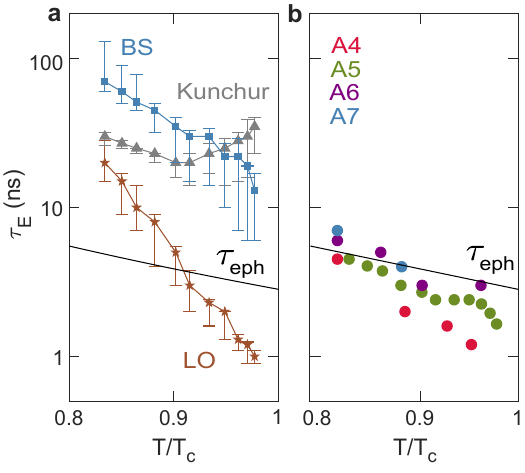}
    \caption{Energy relaxation times as a function of temperature. 
    Energy e-ph relaxation time for single-crystalline TiN films obtained via noise thermometry and spectroscopy \cite{Baeva2024_noise} (black solid line) in comparison with $\tau_E$ obtained from vortex velocimetry. Vortex velocimetry measurements using: LO, BS, and Kunchur models (sample A5; panel a) and modified Kunchur model (samples A3-A7; panel b).}
    \label{figure_num_5}
\end{figure}

Analysis of FFI provides crucial information about energy relaxation processes in vortex cores. Firstly, the critical velocity $v^*$ occurs relatively low in TiN, below 1 km/s, compared to other materials, such as NbN \cite{Haberkorn2024}, NbC \cite{Dobrovolskiy2020_NBC}, MoSi \cite{Budinsk2022}. Given the exceptionally slow e-ph relaxation in this material, a low value of $v^*$ is expected, as it scales inversely with the square root of $\tau_{eph}$. This slow e-ph relaxation in TiN is also beneficial for long recombination times, making it an ideal choice for microwave kinetic inductance detectors \cite{Leduc2010}. Secondly, analysis of $v^*$ by LO, Kunchur, and BS models yields fundamentally different temperature dependencies of $\tau_{E}(T)$. As shown in \autoref{figure_num_5}(a), the LO-derived $\tau_{E}(T)$ does not match the expected $T$-trend of $\tau_{eph}$, exhibiting a two-order-of-magnitude variation across 0.8$T_c-T_c$. Notably, the LO model predicts an unphysical increase in $\tau_E(T)$ with decreasing temperature, which is also reported in other studies \cite{Haberkorn2024,Zhang2020,Leo2011,Ulacco2025}. The BS model also fails to match $\tau_{eph}$ and shows large uncertainties. The Kunchur model systematically overestimates $\tau_{eph}$ while displaying non-monotonic temperature behavior. Although the three existing models predict different behavior for $\tau_{E}(T)$ compared to the expected $\tau_{eph}(T)$, the introduction of coefficients $\alpha$ and $\beta(T)$ into the Kunchur model results in $\tau_{E}(T)$ values that are in agreement with $\tau_{eph}$ in terms of both magnitude and temperature dependence (see \autoref{figure_num_5}(b)). These coefficients account for the energy relaxation path for quaiparticles in the sample and the $T$-dependence of the flux-flow resistance, respectively.

Determining which model best fits experimental data raises several questions. Referring to the B-T diagram in \autoref{figure_num_0}(c), our experimental conditions ($T\lesssim T_x$ and $B \ll B_T$) are a region of uncertainty between the limits of the LO model ($T\gg T_x$ and $B \ll B_T$) and the Kunchur model ($T\gg T_x$ and $B \gg B_T$). 
In this region, a non-uniform vortex flow is expected \cite{Grimaldi2010}, which becomes apparent as the strip width decreases. Introducing a coefficient accounting for the fraction of sample coverage by vortex cores into the Kunchur model results in $\tau_E$ values that consistent with the experimentally found $\tau_{eph}$. These findings suggest that energy relaxation in the vortex core is driven by quasiparticle heating and vortex core expansion, and highlight the importance of thermal effects on the FFI dynamics.

\section{CONCLUSION}
In summary, this study provides a systematic analysis of the FFI effects in single-crystalline TiN films with negligible pinning. Various dynamic FFI models are employed to accurately determine the energy relaxation time $\tau_E$ from the critical velocity measurements. A comparison of $\tau_E$ with values obtained using other independent methods reveals that energy relaxation within vortex cores is governed not by the elevation of quasiparticle temperature respect to phonons, rather than by quasiparticle escape from the core. This finding challenges conventional interpretations and highlights the dominant role of thermal effects in FFI dynamics.

\begin{acknowledgments}
This study, including experimentation and data interpretation, was funded by the Russian Science Foundation grant No. 24-72-10105. The authors would like to thank V. M. Shalaev for providing the TiN film, V. S. Khrapai for discussions, AICF of Skoltech for performing the TEM imaging, and the HSE University Basic Research Program for fabrication of samples. 
\end{acknowledgments}

\section*{Appendix A: The power dissipation}
\begin{figure}[h!]
    \centering
    \includegraphics[width=0.48\textwidth]{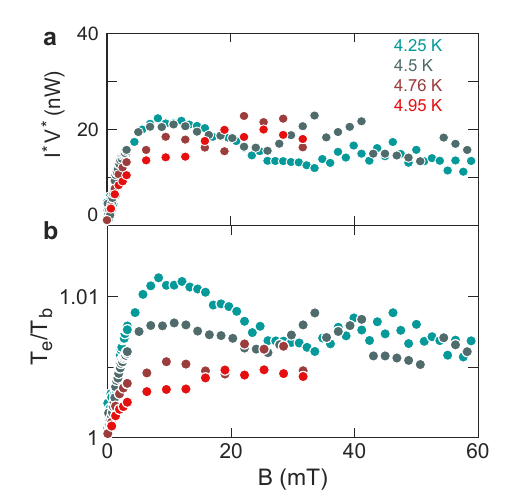}
    \caption{The power dissipation (a) and the ratio of quasiparticle temperature to bath temperature (b) at the FFI point as a function of magnetic field $B$ for sample A5. Different colors represent data measured at different bath temperatures $T_b$ (specified in legend).}
    \label{fig_sup2}
\end{figure}

As shown in \autoref{fig_sup2}, the power dissipation at the flux-flow instability (FFI) point remains relatively low, preventing significant global heating of the system. This conclusion is supported by our calculations of the quasiparticle temperature using the relation $T_e=(T_b+I^*V^*/\nu\Sigma_{eph})^{1/5}$, where the resulting ratio $T_e/T_b$ at the FFI point slightly exceeds 1.01 (a 1\% increase), as shown in \autoref{fig_sup2}(b). This minimal temperature rise confirms that the system remains far from global thermal runaway during FFI.

\section*{Appendix B: Magnetic field dependence of $j_c$}

\begin{figure}
    \centering
    \includegraphics[width=0.48\textwidth]{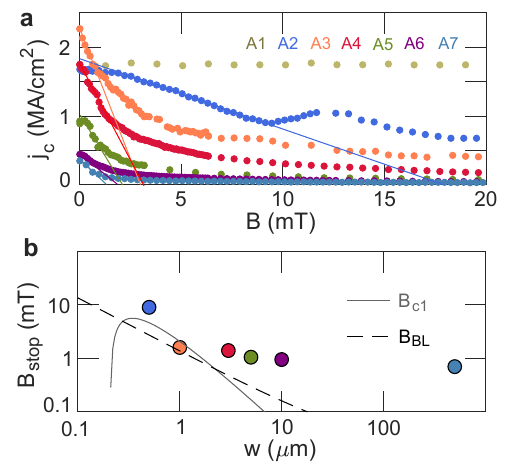}
    \caption{a) Dependence of the critical current density $j_c$ of the samples A1-A7 on the magnetic field on a linear scale. Solid lines correspond to linear fits $j_c(B)=j_c(0)(1-B/2B_{stop})$. b) Width dependence of $B_{stop}$ for the samples A1-A7 (symbols) in comparison with $B_{c1}$ (gray line) and $B_{BL}$ (black dashed line) on a log-log scale.}
    \label{fig_sup1}
\end{figure}
The magnetic field dependence of the critical current density at 4.2 K is presented in \autoref{fig_sup1}(b). At smaller fields, $j_c(B)$ decreases linearly with $B$, while at larger fields the decrease of $j_c$ becomes nonlinear and slower. This behavior can be explained by the presence of some threshold field $B_{stop}$, which demarcates the Meissner (vortex free) and the mixed states of a superconducting strip \cite{Maksimova1998}. Namely, the dependence $j_c(B)$ in the Meissner state ($B<B_{stop}$) is linear and it is described by the expression $j_c(B)=j_c(0)(1-B/2B_{stop})$. Here, $B_{stop}$ may be determined by either the first critical field $B_{c1}=2\Phi_0\ln(w/4\xi_{GL})/\pi w^2$ \cite{Likharev1971} or the field suppressing the surface barrier $B_{BL}=\mu_0 j_s d \sqrt{\Lambda_p/4\pi w+\Lambda_p^2/4w^2}$ \cite{Plourde2001}. \autoref{fig_sup1}(c) compares the measured $B_{stop}$ values with $B_{c1}$ and $B_{BL}$ across different widths. While samples A2-A4 show $B_{stop}\geq B_{BL}$, samples A5-A7 exhibit $B_{stop}$ significantly exceeding $B_{BL}$. This discrepancy can be due to the pinning.

\section*{Appendix C: The flux-flow resistance}
\begin{figure}[h!]
    \centering
    \includegraphics[width=0.48\textwidth]{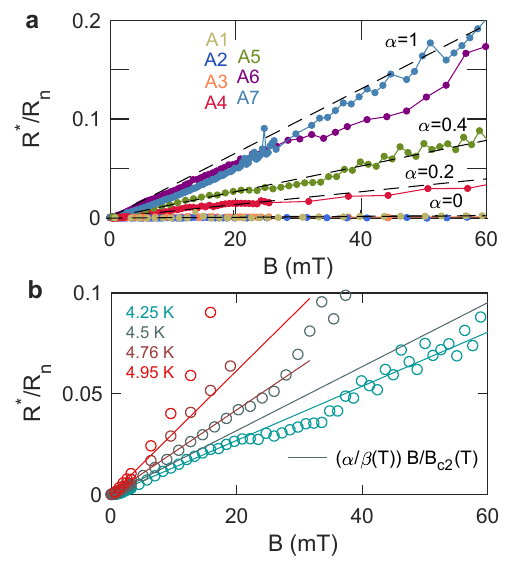}
    \caption{(a) The $B$-dependence of $R^*/R_n$ measured at 4.2K for samples A1-A7 in comparison with $(\alpha/\beta(T)) B/B_{c2}(T)$ function (dashed lines). (b) The $B$-dependence of $R^*/R_n$ for sample A5 in comparison with $(\alpha/\beta(T)) B/B_{c2}(T)$ function with $\alpha=0.4$. Different colors represent data measured at different bath temperatures $T_b$ (specified in legend).}
    \label{fig_sup4}
\end{figure}

The Bardeen-Stephen relation describes the flux-flow resistance at the FFI point ($R^* = V^*/I^*$) in terms of the normal resistance: $R^*/R_n =(1/\beta(T))B/B_{c2}(T)$ \cite{GOR_KOP1}.
In \autoref{fig_sup4}(a), we plot $R^*/R_n$ versus magnetic field for all samples at $T=4.2$ K. The standard Bardeen-Stephen relation (dashed curves) only describes the data well for the two widest samples (A6-A7). For narrower samples, we must introduce a width-dependent coefficient $\alpha$ that accounts for the energy relaxation path for quaiparticles. We observe that $\alpha$ continuously varies from 1 (widest samples) to 0 (narrowest samples), showing a clear width dependence.
 
Using the expression $R^*/R_n =(\alpha/\beta(T))B/B_{c2}(T)$ we can describe the flux-flow resistance at different temperatures. In \autoref{fig_sup4}(b), we plot $R^*/R_n$ versus $B$ for sample A5 and $(\alpha/\beta(T))B/B_{c2}(T)$ where $\alpha=0.4$ and $\beta(T)=1.1/\sqrt{1-T/T_c}$\cite{GOR_KOP1,Gorkov1975}.

\vspace{1cm}

\bibliography{refs}

\end{document}